# Macromolecular crowding modulates folding mechanism of $\alpha/\beta$ protein apoflavodoxin

Dirar Homouz[#], Loren Stagg[+], Pernilla Wittung-Stafshede[+2], and Margaret S. Cheung[#*1]

[#] Department of Physics, University of Houston, Houston, TX, 77204
[+] Department of Biochemistry and Cell Biology, Rice University, Houston, TX 77251, [2]Department of Chemistry, Umeå University, Umeå, 90187 Sweden

---

[1] Corresponding author: mscheung@uh.edu

**Abstract**

Protein dynamics in cells may be different from that in dilute solutions in vitro since the environment in cells is highly concentrated with other macromolecules. This volume exclusion due to macromolecular crowding is predicted to affect both equilibrium and kinetic processes involving protein conformational changes. To quantify macromolecular crowding effects on protein folding mechanisms, here we have investigated the folding energy landscape of an α/β protein, apoflavodoxin, in the presence of inert macromolecular crowding agents using in silico and in vitro approaches. By coarse-grained molecular simulations and topology-based potential interactions, we probed the effects of increased volume fraction of crowding agents ($\phi_c$) as well as of crowding agent geometry (sphere or spherocylinder) at high $\phi_c$. Parallel kinetic folding experiments with purified *Desulfovibro desulfuricans* apoflavodoxin in vitro were performed in the presence of Ficoll (sphere) and Dextran (spherocylinder) synthetic crowding agents. In conclusion, we have identified in silico crowding conditions that best enhance protein stability and discovered that upon manipulation of the crowding conditions, folding routes experiencing topological frustrations can be either enhanced or relieved. The test-tube experiments confirmed that apoflavodoxin's time-resolved folding path is modulated by crowding agent geometry. We propose that macromolecular crowding effects may be a tool for manipulation of protein folding and function in living cells.

# Introduction

Proteins play important roles in the regulation of functions in cells where macromolecules such as nucleic acids, lipids, proteins, and other cytoskeletons can take up to 40% [1-3] of the total cellular volume. In such "crowded" or "concentrated" environments the dynamics of proteins will likely be different from that in dilute solutions in test tubes. Proteins in cell environments experience volume restrictions due to the surrounding macromolecules; this will restrict allowed protein conformations. To circumvent difficulties in experimenting on living matters, experimentalists have used bovine serum albumin (BSA) or synthetic polymers (e.g. Ficoll, Dextran and polyethylglycol, PEG) as "crowding agents" to mimic the crowded conditions of the cell[4]. Using this framework, analyses of crowding effects on protein folding[5-7] and stability[8, 9], protein association[10-12], and protein aggregation[13] have been described.

Nevertheless, it is a challenge to perform such experiments because proteins often aggregate at highly crowded conditions (~200-400 mg/ml). In response to the demand for understanding macromolecular crowding effects, theories and computer simulations are employed to provide mechanistic explanations for these phenomena[14-19]. The most accepted view about crowding effects is that native-state protein stability is enhanced due to compression of unfolded state conformations, attributed to depletion-induced attractions[18, 20, 21]. This idea assumes that perturbation of the native state of the protein is negligible in the presence of crowding. However, recent theories and in vitro experiments suggest that, as protein sizes increase with respect to crowding agents, structural changes in native-states of proteins can occur at high volume fraction of crowding agents[9, 22-24]. However, how protein transient conformations and folding mechanisms are affected by macromolecular crowding is still unclear.

In a recent experimental and computational collaboration on the folding dynamics of apoflavodoxin[9] we revealed that in the presence of high volume fractions of crowding agents stability enhancement of the native protein was in part due to mechanistic protein-crowding agent interactions in the folded state. The crowding agent was modeled by inert hard-core spheres in silico to match the in vitro experiments with Ficoll 70 (a cross-linked, spherical sucrose-based polymer[25]). The native-state effect took place in combination with the generally-accepted destabilization of the denatured state in the presence of crowding agents. *D. desulfuricans* apoflavodoxin is a 148-residue protein containing a central β-sheet surrounded by α-helices on both sides [26, 27]. In vitro far-UV circular dichroism (CD) measurements indicated that the helical content of folded apoflavodoxin was increased by additions of Ficoll 70[9].

Motivated by this study, here we pursue a quantitative, combined in silico and in vitro analysis of the folding mechanism of apoflavodoxin in different crowding conditions varied by both volume fraction ($\phi_c$) and geometry of crowding agents. We find that folding routes are sensitive to the space available to protein dynamics, particularly at high $\phi_c$. At high $\phi_c$, the average void formed by the density fluctuation of the (spherical) crowding agents is small; at this condition elongated, rod-like unfolded ensemble structures are favored that lead to topological frustration when diagnosed by folding route analysis. However, upon changing the geometry of the crowding agent from spherical to anisotropic, the elastic deformation of proteins due to the crowding agents is relieved. This idea can be rationalized by analyzing the space available to a protein, $(1- \phi_c)/\rho_c$, where $\rho_c$ is $N_c/V$ (V is the total volume and $N_c$ is the number of crowding agents larger than solvent molecules.) If we thread two spherical crowding agents into an elongated spherocylindrical (dumbbell) one, while keeping $\phi_c$ the same, the value of $(1-\phi_c)/\rho_c$ doubles as $N_c$ is reduced in half within the same V. As a consequence, the average void volume will become larger for non-spherical crowding agents.

With our in silico model of apoflavodoxin, we show that correct contact formation around the third β-strand in the central sheet is crucial in order to continue folding to the native state, in agreement with previous experimental findings[28]. Upon addition of spherical crowding agents (corresponding to Ficoll 70), we observe an off-pathway folding route that favors early formation of the first terminal β-strand that dominates at high $\phi_c$. This causes topological frustration and the protein must unfold in order to allow the third β-strand to recruit long-range contacts to complete the central β-sheet. Surprisingly, when the spherical crowding agents are replaced by dumbbell-shaped ones, the topological frustration in apoflavodoxin's folding routes vanishes. In agreement, stopped-flow mixing experiments with purified apoflavodoxin in vitro show that folding in buffer and in Ficoll 70 involves rapid formation of an intermediate with ~30 % native-like secondary structure, that is followed by a slow final folding phase; in contrast, in Dextran 70, which is a rod-like crowding agent, apoflavodoxin's intermediate includes ~70 % native-like secondary structure.

# Methods
**SIMULATIONS:**
*Coarse-grained models for proteins and crowding agents*

A coarse-grained sidechain-$C_\alpha$ model (SCM) [18, 29] is used to represent protein structures. Each amino acid (except glycine) is modeled by two beads: a $C_\alpha$ bead from protein backbones and a side chain bead derived from the center of mass of a side chain. An all-atomistic (cartoon) and a coarse-grained SCM representation of apoflavodoxin (using pdb file 2fx2, *D vulgaris* apoflavodoxin) are shown in Figure 1 A and B, respectively. Spherical crowding agents are modeled for Ficoll 70 using hard-core spheres with a radius of 55 Å. Spherocylindrical crowding agents (dumbbell) are modeled for Dextran 70 using two hard-core Ficoll 70 spheres linked by a bond. The distance between two centers of mass is the diameter of a Ficoll 70 (=110Å). By doing so, $N_c$ is reduced in half while others are kept intact. Another small dumbbell composed of two smaller spheres is also used (the length of the bond distance is 87Å) and its volume is equivalent to one Ficoll 70. The potential energy of an entire system with a protein and crowding agents is $E_p+E_{pc}+E_{cc}$. $E_p$ is the potential energy for a protein. $E_{pc}$ and $E_{cc}$ are interactions for protein-crowding agents and crowding agents-crowding agents, respectively. $E_p=E_{structural}+E_{nonbonded}$ where $E_{structural}$ is the structural Hamiltonian and $E_{nonbonded}$ is the nonbonded interactions in a protein. The structural energy, $E_{structural}$, is the sum of the bond-length potential, the bond-angle potential, the dihedral potential, and the chiral potential. Descriptions of each term are provided in previous studies [30]. $E_{non\_bonded}$ includes interactions between a pair of side chains that follows Lennard-Jones potentials ($E^{NB}_{ij}$) and backbone hydrogen bonding that takes the angular alignment of two interacting backbones into account. Nonbonded interactions $E^{NB}_{ij}$ between a pair of *i* and *j* side chain beads at a distance *r* are as follows,

$$E_{ij}^{NB} = \varepsilon_{ij}[(\frac{\sigma_{ij}}{r_{ij}})^{12} - 2(\frac{\sigma_{ij}}{r_{ij}})^{6}] \qquad \text{eqn.1}$$

where $\sigma_{ij} = f(\sigma_i + \sigma_j)$. $\sigma_i$ and $\sigma_j$ are the Van der Waals (VdW) radii of side chain beads. To avoid volume clash, f = 0.9 and $|i-j|>2$.

A modified Go-like model [31] is used in which only native interactions are retained attractive. Consideration for native contact pairs is determined by the CSU program[32]. The amplitude of Go interactions ($\varepsilon_{ij}$) depends on residue types of *i* and *j* and are dictated by the Betancourt-Thirumalai [33] statistical potential. As for non-native interactions, only the repulsive term is used and details of this model are provided elsewhere[30].

For backbone hydrogen bonding interactions, an angular-dependent function that captures directional properties of backbone hydrogen bonds is used in [30],

$$E_{ij}^{HB} = A(\rho)E_{ij}^{NB} \qquad \text{eqn.2}$$

$$A(\rho) = \frac{1}{[1+(1-\cos^2\rho)(1-\frac{\cos\rho}{\cos\rho_a})]^2} \qquad \text{eqn.3}$$

where $E^{NB}_{ij}$ shares the same formula as eqn (1), except that $\varepsilon_{ij}$ for backbone hydrogen bonding is 0.6 kcal/mol and $\sigma_{ij}$ is the hydrogen bond length, 4.6 Å. $A(\rho)$ measures the structural alignment of two interacting strands. $\rho$ is the pseudo dihedral angle between two interacting strands of backbones and it is defined in [30]. $A(\rho)=1$ if the alignment points to β-strands or α-helices. $\rho_a$ is the pseudo dihedral angle of a canonical helical turn, 0.466 (rad). A native pair of backbone hydrogen bonding between $C_\alpha$ and $C_\alpha$ is determined by the DSSP program [34]. Details for this model are provided elsewhere[30].

Nonbonded pairwise interactions between $C_\alpha$ and side chain beads, a protein and crowding agents ($E_{pc}$), and between crowding agents themselves ($E_{cc}$) are repulsive. The repulsion, $E^{rep}_{ij}$, between two interacting beads *i* and *j* at a distance *r*, is as follows:

$$E_{ij}^{rep} = \varepsilon(\frac{\sigma_{ij}}{r_{ij}})^{12} \qquad \text{eqn.4.}$$

where $\sigma_{ij} = \sigma_i + \sigma_j$, $\sigma_i$ and $\sigma_j$ are the VdW radii of interacting beads and $\varepsilon=0.6$ kcal/mol [18].

*Simulation details*

Apoflavodoxin (using pdb file 2fx2) is positioned in a periodic cubic box that includes crowding agents modeled after Ficoll 70 or spherocylindrical crowding agents such as Dextran 70. The box size is twice the length of an extended protein, 300σ, where σ =3.8Å is the average distance between two adjacent $C_\alpha$'s. The volume fraction of crowding agents, $\phi_c$, is $4\pi N_c R_c^3/3V$, where $R_c$ is the radius of a Ficoll 70 (55Å), $N_c$ is the number of crowding agents, and V is the total volume. $\phi_c$= 0 (bulk), 25%, and 40%, are chosen as it represents a broad range of volume fractions in a cell.

Thermodynamic properties are obtained by molecular simulations where the Langevin equations of motion at a low friction limit are implemented [35].To enhance sampling efficiency in a simulation, a replica exchange method (REM) is used [36] to incorporate high-performance computing resources nationwide. Details of REM simulations can be found in previous studies[9, 18]. Temperatures for REM simulations range from 250 to 450K and more than 40 temperatures (or replica) within this range are assigned. Starting configurations are randomly taken from high-temperature simulations and then quenched to starting temperatures. The integration time step is $10^{-4}\tau_L$, $\tau_L = (m\sigma^2/\varepsilon)^{1/2}$, m is the mass of a residue[35]. At certain steps ($40\tau_L$), configurations with "neighboring" temperatures are exchanged. The exchange rate between each replica is ~40%. A total number of 40000 statistically significant data points is collected from each replica for free energy computations. For a bulk case at T=360K, there are about 40 folding/unfolding transition events within 40,000 data points. The weighted histogram analysis is employed to extrapolate thermodynamic properties and to compute errors[37].

*Clustering analysis on ensemble structures*

Ensemble structures are dissected into the folded state, the unfolded state, and the transition state by their locations in a folding energy landscape. A protein folding energy landscape is plotted as function of Q (the fraction of native contacts where Q ranges from 0 to 1 and Q=1 is the native state) and it presents with double minima separated by a barrier at Q*. The region of the folded state is chosen within ±20% of the minimum near Q=1. For the unfolded state, configurations within ±20% of the other free energy minimum near Q=0 are collected. For the transition state, configurations within 20% of the peak of a barrier at Q* are collected.

Configurations in each region are sorted into fewer sets of clusters using a clustering method [38, 39] based on a self-organized neural net algorithm[40, 41] which does not require a prior structural knowledge of an ensemble. A conformation of $j$ is described by a vector $x_j$. Elements of $x_j$ are the separations between pairwise side chain beads in j. A Euclidean distance between the *j*th conformation and the center of mass of the *k*th cluster is measured and compared to a cutoff. Conformations of *j* will be individually sorted to the *k*th cluster if this Euclidean distance is shorter than a cutoff. After all conformations are sorted, the center of mass of the *k*th cluster will be re-evaluated and the sorting of each conformation will start over again. This process is performed iteratively until the centers of masses of all clusters converge. Details of the procedure are given in [38]. The cutoff in this work is about ~60$\sigma$, depending on the number of clusters generated (~10 clusters).

*Shape analysis*

The shape of configurations can be characterized by two rotationally invariant quantities, the asphericity ($\Delta$) and shape parameter (S)[42]. Asphericity ($\Delta$) ranges from 0 to 1 and $\Delta$=0 corresponds to a sphere. Deviation of $\Delta$ from 0 indicates an extent of anisotropy. The shape parameter (S) ranges from -0.25 to 2. S<0 corresponds to oblates and S>0 to prolates, while S=0 is a sphere.

**EXPERIMENTS**

*Protein preparation*

Apoflavodoxin from *D. desulfuricans* (ATCC 29577) was expressed in *Escherichia coli* cells and purified as previously described[43] with some modification. In short, the protein was first isolated using a Q-Sepharose column and further purified using a Superdex-75 gel filtration column using an AKTA FPLC system (Amersham-Amersham Pharmacia).

*In Vitro Measurements*

Thermal unfolding experiments with apoflavodoxin were performed using CD (Jasco-810 instrument) in 10 mM HEPES (pH 7). CD was monitored at 222 nm from 20 to 95 °C at a rate of 2.5 °C/min. No scan rate dependence was found between 0.1 °C/min and 2.5 °C/min. Temperature was controlled with a Jasco PTC 424S peltier. Separate experiments were carried out with 100 mg/ml increments of Dextran 70 (Amersham Biosciences) between 0 and 400 mg/ml. Urea-induced equilibrium unfolding of apoflavodoxin was monitored by CD in the presence of final concentrations of 0, 75, and 150 mg/ml Dextran 70 and Ficoll 70 (Amersham Biosciences) in 10 mM HEPES (pH 7) at 20 °C. High quality urea (Sigma-Aldrich) was made fresh to a stock of 10 M and filtered using 0.45 μm syringe filters (Fisher) before use. Each CD spectrum (260-200 nm) was a result of averaging two successive scans. Samples were made at 0.25 M increments of urea between 0 and 6 M and allowed to equilibrate at room temperature for approximately 1.5 hrs prior to measurements. Urea-induced kinetic folding/unfolding of apoflavodoxin was monitored using CD at 222 nm with an Applied Photophysics Pi-Star stopped-flow mixer as a function of Dextran 70 and Ficoll 70 (0, 75, and 150 mg/ml). In the case of Ficoll 70 which is assumed to be spherical, calculations suggest that 75 mg/ml should correspond to ~35 % volume occupancy; thus 150 mg/ml Ficoll 70 would represent ~70 % volume occupancy. However, these values are likely overestimates as it is possible to make Ficoll solutions as concentrated as 400 mg/ml (although the shape may not be spherical at the higher mg/ml:s). Temperature was maintained at 20 °C using a Julabo F30-C peltier; buffer was 10 mM HEPES (pH 7). Apoflavodoxin was unfolded to proper concentrations of urea using 1:5 mixing and refolded by dilution upon 1:10 mixing. Stock protein solutions were equilibrated at room temperature for 1.5 hrs before measurement. Observed kinetic traces were fit to single exponential decays. Missing amplitude in the mixing step was estimated from the baselines of folded and unfolded protein signals. All experiments were carried out with a final protein concentration of 20 μM.

# Results

*Macromolecular crowding enhances protein stability*

We find that folding temperatures ($T_f$, defined by the temperature for $\Delta G_{fu}=0$) for apoflavodoxin unfolding increase with $\phi_c$. In the bulk case (i.e., $\phi_c =0$) $T_f$ is 354 K. $\phi_c$ (Ficoll 70)=25%, $T_f$=360K and $\phi_c$ (Ficoll 70)=40%, $T_f$=373 K (also from [9]). This indicates enhancement of protein stability in the presence of crowding agents. Interestingly, when dumbbell-like crowding agents are included, despite the same volume fraction, $T_f$ increases ($\phi_c$ (dumbbell)=40% and $T_f$=388 K) more than that in the presence of the spherical crowding agents. This is also found experimentally in heating experiments with purified apoflavodoxin detected by far-UV CD changes. Thermal midpoints for apoflavodoxin in various amounts of Ficoll 70 have been reported previously [9]. We here performed similar experiments in the presence of the same amounts of Dextran 70 and compared the thermal stability of apoflavodoxin in the two crowding agents. We find that the effects of Dextran 70 at each condition are larger than with Ficoll 70 (Table 1).

Enhancement of protein stability due to crowding can be also observed in the free energy profiles plotted as a function of Q at different crowding conditions (Figure 2). The stability enhancement can be explained by density fluctuation of crowding agents; these prompt the protein to reside at low density regions so that compact unfolded protein conformations are favored on average (The radius of gyration, $R_g$, of unfolded proteins, reduced in the presence of crowding agents. See Table S1 in the supplement materials). This depletion-induced attraction is more prominent to the ensemble of unfolded states than in the folded states. While destabilization of unfolded states by crowding is much discussed [20], how crowding agents affect folded states of large proteins (such as apoflavodoxin) is poorly understood. At $\phi_c$=25 %, the free energy minimum of apoflavodoxin's folded state ($Q_f$) shifts towards 1 (Q=1 is the crystal structure)[9]. This indicates that the folded state resembles the crystal structure more in the presence of crowding agents than in buffer, while native-state shifts in Q have not been observed for small proteins and peptides[18]. It can be explained by mechanistic interactions between a large, malleable apoflavodoxin and the crowding agents resulting in more tightly formed contacts in frayed loops, floppy termini regions, and those between sheets and helices.

Nevertheless, at higher $\phi_c$ (e.g. $\phi_c$(Ficoll 70)≥ 25%), mechanistic interactions by the crowding agents may cause local deformation of the folded state (protein is modeled as elastic object). This may compromise the favorable effects due to crowding on protein stability. To dissect such subtle differences, we performed composite analyses of energy and entropy using the energy landscape theory (ELT) framework [44-46]. Differences in energy ($\Delta E_{fu}$) and entropy ($\Delta S_{fu}$) between the folded and the unfolded states in the presence of crowding agents are given in Table 2. At low $\phi_c$ (e.g. $\phi_c$ < 25 %), because the crowding agents act on the protein with repulsive forces, there is a slight reduction in energy ($\Delta E_{fu}$) compared to the bulk case ($\phi_c$ =0). Nevertheless, as crowding effects produce more compact unfolded states, the difference in entropy between the folded and unfolded states ($\Delta S_{fu}$) further decreases.

At $\phi_c$ =40%, nonspecific repulsive interactions between crowding agent and the folded protein play predominant roles in apoflavodoxin stability. Elastic deformation is non-negligible and it offsets the energy bias ($\Delta E_{fu}$). However, the folded state at high $\phi_c$ is still favored thermodynamically for two entropic reasons. First, the entropy of the unfolded state decreases due to compaction of the unfolded states. Second, entropy of the folded state increases due to a large growth of crowding-induced states that are not originally accessible in the bulk case. As a consequence $\Delta S_{fu}$ between the folded and unfolded are much less than $\Delta E_{fu}$ and therefore the free energy stabilization ($\Delta G_{fu}$) at $\phi_c$ =40 % strongly favors the folded state.

We also analyzed the free energy landscape for this condition when spherical crowding agents are replaced by dumbbell crowding agents (Figure 2). At the same volume fraction ($\phi_c$ =40 %), the minimum of the unfolded state ($Q_u$) with dumbbell crowding agents shifts to $Q_u$=0.38, compared to $Q_u$=0.30 in presence of spherical crowding agents. This indicates that more native contacts are formed in the unfolded state in the presence of the dumbbell crowding agents than in the presence of the spherical ones. Moreover, the stability of unfolded states in the presence of dumbbell crowding agents decreases such that the stability of the folded state ($\Delta G_{fu}$) is relatively favored.

We speculate this enhancement in stability at $\phi_c$(dumbbell)=40% is partly due to better chances of forming larger voids that are available to the protein as two spherical crowding agents are linked into one in a pool of fluctuating crowding agents. Despite $\phi_c$ remaining the same, the space available to a protein, $(1-\phi_c)/\rho_c$, doubles in the dumbbell conditions as the number of dumbbell particles, $N_c$(dumbbell), is half of Ficoll 70, $N_c$(Ficoll 70). In the larger voids formed upon fluctuations of dumbbell crowding agents, there is less crowding-induced deformation in the folded state at high $\phi_c$ and the distribution of structures in the unfolded state may also be altered accordingly. This speculation can be partly supported by the analysis of $\Delta E_{fu}$ and $\Delta S_{fu}$. The differences in both energy and entropy are larger with dumbbell crowding agents than with Ficoll 70 (Table 2). It is tempting to suggest this type of adjustment of the space available to a protein at high $\phi_c$ may result in changes in protein folding mechanisms. In the next section we explore the folding mechanism of apoflavodoxin at different crowding conditions.

*Crowding effects on the folding mechanism of apoflavodoxin*

The folding mechanism of apoflavodoxin was studied upon analysis of the behavior of selected groups of native contacts in the evolution of Q on the folding energy landscape. This profile will measure the heterogeneity of folding routes by following contact formation from specific regions. For example, if all contacts are formed in a homogeneous fashion, the behavior of the *i*th group of contacts will be the same as an average behavior from the whole. Such a mean-field like behavior will show that the probability of contact formation in the *i*th group ($<Q>_i$) progresses linearly with Q. However, if heterogeneity in contact formation exists, the rise and fall of route profiles, presented by $<Q>_i$ *vs.* Q, will be diagnostic of topological frustration in a funnel-like folding energy landscape. Several studies have used this and other similar parameters[47] to evaluate folding routes of protein L[48] and β-trefoil proteins[49].

Using this diagnostic tool, we monitored contact formations in the first β-strand ($β_1$), the first α-helix ($α_1$), and the third β-strand ($β_3$) regions. These places are identified as particularly interesting in apoflavodoxin. $<Q>_i$ for these regions (*i*: $β_1$, $α_1$, and $β_3$) is plotted as a function of Q in Figure 3. In the absence of crowding agents (Figure 3A), $<Q>_{α1}$ rises sharply when Q<0.2, indicating that contacts in $α_1$ (red) form early in the unfolded states. In the TSE region (Q~0.5), $<Q>_{β3}$ rises quickly as compared to the rest, as contacts in $β_3$ (green) form predominantly and the major core in the central β-strand region is being constructed. Contacts in $β_1$ (black) are greatly suppressed until the very late stage of folding (Q>0.7). These folding pathways that involve recruiting contacts in $β_3$ in the TSE are agreeable with experimental data[50] and simulations[51] on folding of apoflavodoxin.

At high levels of crowding agents ($\phi_c$(Ficoll 70)=40%) in Figure 3B, depletion-induced attraction from the presence of Ficoll 70 results in early contact formation in terminal $β_1$ prior to central $β_3$ as $<Q>_{β1}$ rises sharply at Q=0.4. Surprisingly, to further continue folding to the folded state, contacts in $β_1$ have to unfold (as $<Q>_{β1}$ dramatically drops from ~0.4 to 0.3) at the TS (Q*=0.55) in order to make clear ways for $β_3$ to recruit other long-range contacts and form the β-sheet core. This situation clearly elucidates topological frustration in the folding of apoflavodoxin as early contact formation in $β_1$, despite being native ones, hinders further folding. We analyzed the conformations causing topological frustration and found that there is a small loop formed in the terminal $\beta_1$ region (Figure 3D).

However, as we consider the folding routes for apoflavodoxin with dumbbell crowding agents at $\phi_c$ (dumbbell)=40 % in Figure 3C, the pattern changes dramatically. The profile for $<Q>_{\beta_1}$ along Q in Figure 3C is significantly more coherent with $<Q>_{\beta_3}$ than those in Figure 3A and 3B, indicating that contact formations in $\beta_1$ and $\beta_3$ come together in a homogeneous fashion. We speculate that the distribution of unfolded state conformations, which depend on crowding agent geometry, dictates partitioning between apoflavodoxin folding routes.

To assess this finding by in vitro experiments, we performed a set of time-resolved mixing experiments with apoflavodoxin in the presence of Ficoll 70 and Dextran 70 at 20°C (pH 7) using urea as the denaturant. Apoflavodoxin unfolds in an apparent two-state reaction upon urea additions (inset Figure 4A). The presence of 75 and 150 mg/ml Ficoll 70 gradually increases the unfolding midpoint and the unfolding-free energy (obtained from two-state fits to the data). The same trends are observed with Dextran although the effects are somewhat larger (inset, Figure 4B). Unfolding and refolding kinetics were probed by urea jumps to conditions favoring unfolded and folded states, respectively in 0, 75, and 150 mg/ml Ficoll 70 and Dextran 70. The semi-logarithmic Chevron plots for apoflavodoxin are shown in Figure 4A (Ficoll 70) and 4B (Dextran 70). In buffer, apoflavodoxin folds via a burst phase intermediate with about 30 % of the total native-state CD signal at 222 nm, which reports on secondary structure content. This phase is followed by a slow exponential conversion of the intermediate to the native state [27]. The same kinetic mechanism is found in the presence of Ficoll 70. Like in buffer there is an initial burst intermediate (< ms) with about 30 % of the native CD signal. In agreement with predicted crowding effects [18], the subsequent final folding event is 2-3 fold faster in the presence of Ficoll 70 than in buffer. When the same refolding experiments were repeated with apoflavodoxin in the presence of Dextran 70, we found that the burst phase intermediate now included 70 % of the native state's secondary structure content. Thus, after a few ms of folding, the structural content in apoflavodoxin is more than 2-fold higher in Dextran than in Ficoll. The subsequent final folding step of apoflavodoxin in Dextran was similar in terms of rate constant as in Ficoll 70. These experimental results support the in silico findings of different folding routes in the different crowding agents. Although there is an intermediate in both cases, the structural content of the intermediate differs dramatically between the two crowding agents (although we cannot assess experimentally if any of these structures are misfolded or not). Rate constants for unfolding and the final folding step at each condition, extrapolated to zero denaturant concentration, are reported in Table 3.

*Clustering analysis on ensemble structures*

To explain deviations of folding routes at different crowding conditions, we further dissect the characteristics of each state at T=360K using clustering analysis (see the method section for details).

<u>Unfolded states</u>: In Table 4, the percentage of unfolded conformations in the first dominant cluster under different crowding conditions is provided. As $\phi_c$ (Ficoll 70) increases to 40%, the size of the first dominant cluster increases by 2.6 fold, compared to the bulk. Interestingly, when dumbbell-like crowding agents are considered ($\phi_c$ (dumbbell)=40%), the size increases by 3.6 fold.

Next, we compute the asphericity ($\Delta$) and the shape (S) of conformations in the first dominant cluster (Table 5). In the bulk case, the ensemble structures in unfolded states are mostly prolate (S=0.448±0.008) and rod-like ($\Delta$=0.389±0.003). Upon addition of Ficoll 70, the geometric characteristics in the dominant cluster of unfolded states are still prolate and rod-like (Table 5). Given elongated, rod-like conformations, loop formation at the terminal $\beta_1$ that causes topological frustration is overwhelmingly favored. This can explain the enhanced topological frustration diagnosed in Ficoll 70 from the folding route analysis in Figure 3.

Interestingly, with dumbbell crowding agents ($\phi_c$ (dumbbell)=40%), the geometric characteristics of the dominant unfolded ensemble becomes spherical (S=0.078±0.011). Given more spherical and compact structures, contact formation in the central $\beta_3$ can better recruit other long-range contacts to form the core $\beta$-sheet as the chances of contact formation in terminal $\beta_1$ diminished. Patterns of folding routes with dumbbell crowding agents is therefore more homogeneous (Figure 3C) than with spherical crowding agents (Figure 3B).

*Folded states:* In the presence of crowding agents depletion-induced attraction produces more compact structures and the size of the first dominant cluster resembling the native crystal state increases. At high $\phi_c$, these ensemble structures are compressed into slightly oblate shapes (S=-0.012±0.000), Table 4. This behavior was discussed further in our previous equilibrium analysis of apoflavodoxin in the presence of Ficoll 70[9].

*Transition states:* Probability of contact formation in the transition state ensemble (definition of the transition state is in the method section) at different crowding conditions $\phi_c$ (Ficoll 70) =0, 25%, 40% and $\phi_c$ (dumbbell) =40% are given in Figure S1 (A) to (D) (See supporting materials). Both axes indicate residue indices. Elements in the upper triangles in Figure S1 are the probability of contact formation between backbone hydrogen bonds and the lower triangular ones are contact formation between side chain beads. Even at high volume fractions of the crowding agents microscopic descriptions of the transition state structures measured by native contact formation are similar to those in the bulk. Short-range, on-diagonal contacts and contacts in the central $\beta$-strands are mostly formed. The contacts in the TSE in the presence of crowding agents are not greatly affected by crowding agents, except that the overall molecules are more compact as seen from the increase in the probability of off-diagonal, long-range contacts.

Despite the similarity in microscopic details in TSE, the overall geometry of the TSE structures is mildly susceptible to protein-crowding interactions at high $\phi_c$. Shape and asphericity of the first dominant clusters are given in Table 4 and the TSE structures are shaped by crowding interactions into spheres. These changes in shapes, however, are not as dramatic as those in the unfolded states at different crowding conditions.

## Discussion

*Volume interactions and the geometry of crowding agents on protein stability*

Volume fractions of crowding agents ($\phi_c$) play a major role in the enhancement of protein stability, as (1-$\phi_c$)/$\rho_c$ defines the range of space available to accommodate proteins. To this end, density fluctuations of crowding agents are also important as they dictate the available ranges for structural fluctuations in protein ensembles, particularly in our case that the crowder size is much bigger than an apoflavodoxin protein. This is of particular importance in our case, since the size of the crowder is much bigger than the protein. It was shown in both experiments and simulations that at low $\phi_c$ (≤25%) or low weight concentration (≤100 mg/ml) the effect of crowding on protein thermodynamics is minimal. The size effect on biopolymer folding was earlier studied by Minton[52] and it was suggested that large crowding agents impose less effect on conformations of small biopolymers as compared to small crowding agents. However, as $\phi_c$ increases, interactions of crowding agents with folded forms of proteins are not negligible and protein dynamics becomes sensitive to the geometry of crowding agent.

Our results show that crowding agents acts to the native state at high $\phi_c$ ($\phi_c$ =40%). To investigate the role of crowding agent geometry in these conditions we consider two extreme cases that deviate from spherical Ficoll 70. The first case was to put two spherical Ficoll 70 molecules together into an elongated dumbbell. This dumbbell has an effective radius of gyration 25 % larger than the Ficoll 70 sphere. Essentially, given the same $\phi_c$, the available space $(1-\phi_c)/\rho_c$ to a protein will be doubled when the number of crowding agents $N_c$ is reduced by half. It gives us a controlled way to address the importance of the void size, caused by fluctuations of crowding agents. Despite a small change, we showed here that the effect of this on protein stability and folding mechanism is significant. We find that at high $\phi_c$, an increase in $(1-\phi_c)/\rho_c$, by introducing different crowding agent geometry, that results in larger voids can relieve mechanistic interactions between crowding agents and proteins.

The other case was to create a dumbbell that has the equivalent volume as Ficoll 70 (we name this "small dumbbell" in the method section). According to Table 2, there is still an enhancement of the native state stability, but the effect is not as significant as by the other agents at $\phi_c$=40%. The size of the depletion layer in the protein-crowder radial distribution function is the smallest for the small dumbbell (which is in line with Berg's analysis on the behavior of dumbbell molecules with equivalent volume of spheres[53]), indicating significant unfavorable protein-crowder interactions that offset the protein-stability enhancement (data not shown). Together with Ficoll 70, these two dumbbell models provide a framework to discuss the characteristics of shape and size of crowding agents. We argue that the best computer model for Dextran 70 may be a "swelled" dumbbell (and not the small dumbbell) as Dextran 70 has lower density than Ficoll 70[25].

It was noted that cylindrical crowding agents can exclude more volume than spherical crowding agents and the protein stability can be more enhanced. This appears supported by some experimental studies. In addition to the experimental data reported here for apoflavodoxin (Table 1), in a recent experimental study, Perham et al. showed that Dextran promoted larger effects than Ficoll 70 on thermal stability of holoflavodoxin and of another unrelated protein, VlsE[23].

Summarizing the numerical analysis results, what is the crowding condition that best promotes protein stability? The dominant factor is no doubt $\phi_c$. However, at high $\phi_c$, protein-crowder interactions may also start bringing negative impacts. Therefore, there appears to be another competing factor, available space to a protein is also important in defining the folding mechanism and protein stability. Space available to proteins is related to the probability of void formation which increases with $(1-\phi_c)/\rho_c$ as seen from previous studies of fluid models[54, 55]. The available space can be adjusted by the geometry of the crowding agents. This was also discussed by Minton in his early study that the shape of crowding agents is important to the protein stability[56]. We suggest that the stability of a native protein, $\Delta G_{fu}$, in the presence of $\rho_c$ crowding agents, is partly proportional to the following equation,

$$\Delta G_{fu} \sim (\alpha \phi_c + \beta \phi_c^2 + \gamma \phi_c^3)/\rho_c \qquad \text{eqn. 5}$$

This equation is based on the fact that $\Delta G_{fu}$ is primarily determined by $\phi_c$ (up to a critical $\phi_c^*$). We propose that there exists $\phi_c^*$ that defines the optimal volume fraction of crowding agents that best enhances protein stability. $\beta$ is a parameter that defines how crowding agent geometry influences the available space of a protein. In this study $\beta$ is likely a negative value, as protein-crowding interactions at high $\phi_c$ provide some unfavorable effects on the stability of the native state. Thus at $\phi_c > \phi_c^*$, a second-order effect is introduced so that the protein native stability is less enhanced. By changing (c (or equivalently Nc) we can manipulate this second-order effect. In addition, our results from the small dumbbell with the equivalent volume of a Ficoll 70 suggested a third factor that arises from the shapes of crowding agents which relates to the extent of packing of a protein and crowders in three-dimensional space. This is a many-body problem that can be included in the third order term ($\gamma$) in the equation above (it will be pursued in future studies). Certainly, there are still other factors to consider by going beyond the hard-core approximation, such as attractive interactions between constituents[57, 58]. This relation based on the excluded volume indicates that by manipulating $\phi_c$ and the geometry of crowding agents, switch-like protein activities (that are triggered by structural changes) may be achieved in a cell.

*Crowding effects on folding mechanism of apoflavdoxin*

Apoflavodoxin is a large protein with a central β-sheet sandwiched by α-helices at both facets. Recent experiments have shown that multiple folding routes exist and some misfolded structures need to unfold in order to continue its route to the folded state[50]. In this process, the formation of the third β-strand is crucial for recruiting long-range contacts and this behavior was modeled to a nucleation-growth model[51].

We investigated crowding effects on these folding routes by profiling regional contact formation along with Q on the folding energy landscape. At high $\phi_c$ of spherical crowders, available spaces for unfolded states are limited and unfolded states adopt prolate shapes. As a consequence, local interactions along the chain are favored and that is found to cause population of contacts in the terminal $\beta_1$. Despite native contacts, their formation causes topological frustration and hinders further contact formation in $\beta_3$. "Topological frustration" in the folding mechanism has also been seen in other protein families[49, 59]. *However, this is the first study that suggests using crowding agent geometry to get around topological frustrations in folding pathways.*

**Conclusions**

The effects of macromolecular crowding on the stability, structures, and folding routes of apoflavodoxin were investigated using a combined approach of molecular simulations, the energy landscape theory and in vitro measurements. Our results show that native protein stability is partially proportional to the volume fraction of crowding agents ($\phi_c$) and the space available to a protein: $(1- \phi_c)/\rho_c$. The latter is significant at high $\phi_c$ and the protein stability can be fine-tuned using different geometry of crowding agents. We found the folding mechanism of apoflavodoxin to be modulated by changes in crowding conditions. Conditions were found that altered the folding route so that topological frustration was avoided. We propose that selective crowding conditions may be a useful tool to manipulate protein-based biological reactions in vitro and in vivo.

**Acknowledgments**

D.H. thanks S.Q. Zhang for his help in preparation steps. M.S.C. thanks the GEAR grant from the University of Houston and the TcSUH seed grant for the financial support. Computing resources are partly supported by the Texas Learning Computing Center and the National Science Foundation through TeraGrid resources provided by TACC and SDSC (MRAC: TG-MCB070066N and MRAC:TG-MCB080027N). PWS acknowledges the Welch Foundation for funding (C-1588).

**Tables:**

**Table 1:**

| Crowding Agent (mg/ml) | Ficoll 70 | Dextran 70 |
|---|---|---|
| 0 | 317 | 317 |
| 100 | 321 | 321 |
| 200 | 325 | 327 |
| 300 | 331 | 337 |
| 400 | 337 | 344 |

Thermal midpoints (in Kelvin) for apoflavodoxin unfolding as detected by far-UV CD in different amounts of Ficoll 70 [9] and Dextran (10 mM Hepes buffer). Error in each value is +/- 1 K.

**Table 2:**

| $\phi_c$ | $\Delta E_{fu}$ | $\Delta S_{fu}$ |
|---|---|---|
| 0% (bulk) | -86.92 | -86.97 |
| 25 % (Ficoll 70) | -83.89 | -83.53 |
| 40 % (Ficoll 70) | -21.86 | -20.03 |
| 40 % (dumbbell) | -23.76 | -20.83 |
| 40% (small dumbbell*) | -22.25 | -20.46 |

Differences (in unit of $k_BT$) in energy ($\Delta E_{fu}$) and entropy ($\Delta S_{fu}$) between the folded and unfolded states at 354K at various crowding conditions. These values are derived from free energy computation in Fig.2. The estimated errors are of the same order as free energy (~0.05$k_BT$). Small dumbbell* is a spherocylindrical crowder with the equivalent volume as a one Ficoll 70.

Table 3:

| Crowding Agent (mg/ml) | $k_f$ (s$^{-1}$) | $k_u$ (s$^{-1}$) |
|---|---|---|
| 0 | 0.39±0.02 | 0.016±0.005 |
| 75 (Ficoll 70) | 0.56±0.02 | 0.016±0.005 |
| 150 (Ficoll 70) | 0.95±0.02 | 0.010±0.005 |
| 75 (Dextran 70) | 0.53±0.02 | 0.014±0.005 |
| 150 (Dextran 70) | 0.95±0.02 | 0.009±0.005 |

Rate constants for the final folding step (after the burst) and unfolding for each condition (extrapolated to 0 M urea) extracted from the data in Figure 4.

**Table 4:**

| φc | Cluster size of the folded state | Cluster size of the unfolded states |
|---|---|---|
| 0% (bulk) | 63% | 10% |
| 25 % (Ficoll 70) | 90% | 16% |
| 40 % (Ficoll 70) | 99% | 26% |
| 40 % (dumbbell) | 100% | 36% |

Clustering analyses of ensemble structures. The cluster size is the percentage of conformations in the dominant cluster.

**Table 5:**

Folded state

| $\phi_c$ | Δ | S |
|---|---|---|
| 0% (bulk) | 0.037±0.000 | -0.012±0.000 |
| 25 % (Ficoll 70) | 0.036 ±0.000 | -0.012±0.000 |
| 40 % (Ficoll 70) | 0.036±0.000 | -0.012±0.000 |
| 40 % (dumbbell) | 0.036±0.000 | -0.011±0.000 |

Transition state

| $\phi_c$ | Δ | S |
|---|---|---|
| 0% (bulk) | 0.029±0.000 | -0.001±0.000 |
| 25% (Ficoll 70) | 0.035 ±0.000 | 0.001±0.000 |
| 40% (Ficoll 70) | 0.031±0.000 | -0.008±0.000 |
| 40% (dumbbell) | 0.037±0.000 | -0.004±0.000 |

Unfolded state

| $\phi_c$ | Δ | S |
|---|---|---|
| 0% (bulk) | 0.389±0.003 | 0.448±0.008 |
| 25% (Ficoll 70) | 0.265 ±0.006 | 0.212±0.013 |
| 40% (Ficoll 70) | 0.306±0.009 | 0.276±0.019 |
| 40% (dumbbell) | 0.138±0.008 | 0.078±0.011 |

Asphericity(Δ) and shape(S) parameters of structures from the dominant cluster at different crowding conditions.

**Figure Legends**

Figure 1: An $\alpha/\beta$ apoflavodoxin protein in (A) an all-atomistic cartoon representation and (B) a coarse-grained side chain-C$_\alpha$ model (SCM) representation. Structures are colored by index numbers.

Figure 2: Free energy profiles are plotted as a function of Q (the fraction of native contact formation) at different crowding conditions at 354K. $\phi_c$ (bulk)=0, solid line; $\phi_c$ (Ficoll 70)=25%, dotted line; $\phi_c$ (Ficoll 70)=40%, dashed line; and $\phi_c$ (dumbbell)=40%, dot-dashed line. Error bars are included.

Figure 3: Probability of select native contact formation <Q>$_i$ at the *i*th region of a protein in the evolution of protein folding. Contact formation of the first β-strand (black), the first α-helix (red), and the third β-strand (green) is plotted as a function of Q in (A) bulk, (B) $\phi_c$ =40%, Ficoll 70, and (C) $\phi_c$ =40%, dumbbell crowding agent, respectively. (D) shows a typical conformation in the unfolded state with some contacts formed about β$_1$ in early Q that cause topological frustrations in the folding landscape. The diagonal line (blue) is provided as a visual guidance for a mean-field like behavior. Error bars are included.

Figure 4: Kinetic semi-logarithmic plots of lnk versus [urea] for apoflavodoxin folding/unfolding in buffer (filled circles), 75 mg/ml crowding agent (open squares) and 150 mg/ml crowding agent (filled triangles) using Ficoll 70 (A) and Dextran 70 (B). The kinetic folding mechanism involves a burst phase taking place within the mixing time (i.e., in < ms) which corresponds to 30 % of the total CD change in buffer and in Ficoll, but 75 % of the total CD change in Dextran. The measurable rate constants for the subsequent folding of the intermediate to the native state are shown. In both crowding agents, the rate constants extrapolated to 0 M urea increases 2-3 fold as compared to in buffer alone (Table 3). There is no significant effect on the unfolding rates extrapolated to 0 M urea in the different conditions (Table 3). The insets show the equilibrium unfolding curve as a function of [urea] for each condition. Two-state fits to the equilibrium data reveal that the free energy of unfolding increases from 4.4 kJ/mol (buffer) to 8.9 kJ/mol (150 mg/ml Dextran) and to 7.2 kJ/mol (150 mg/ml Ficoll 70).

1:

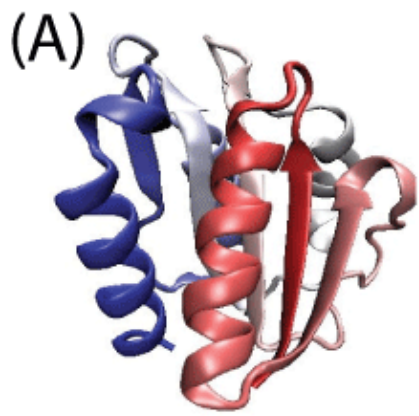 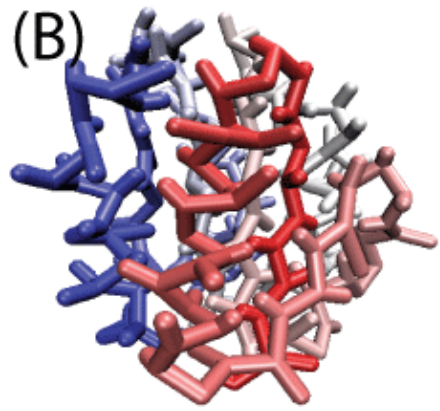

2:

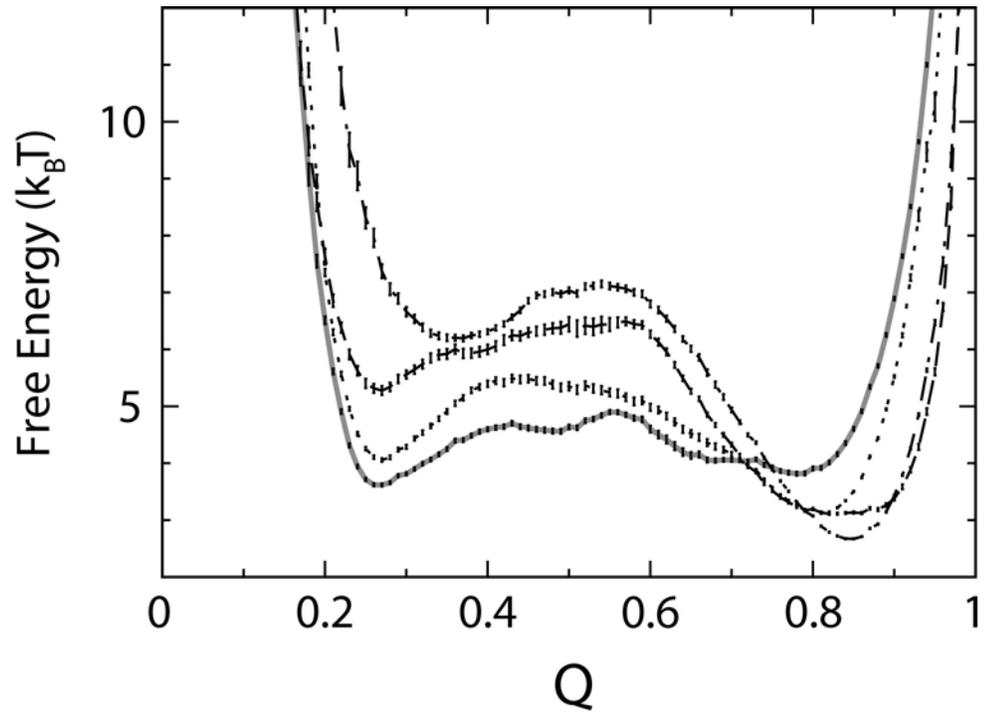

3:

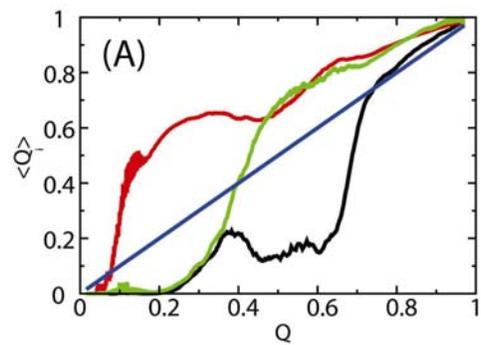 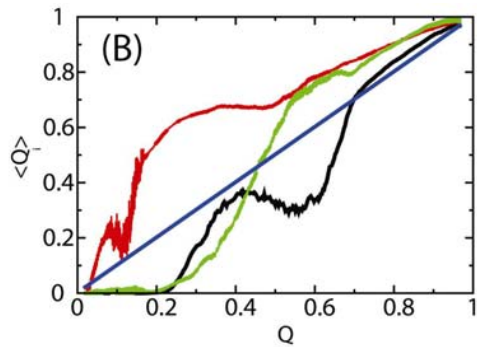

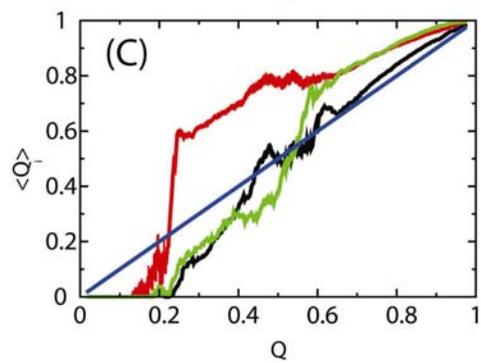 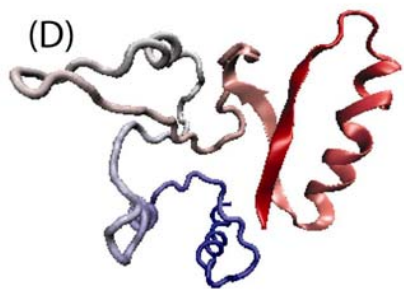

**4:**

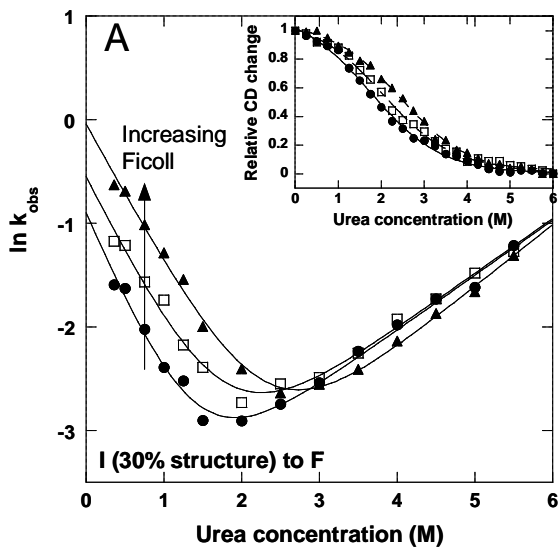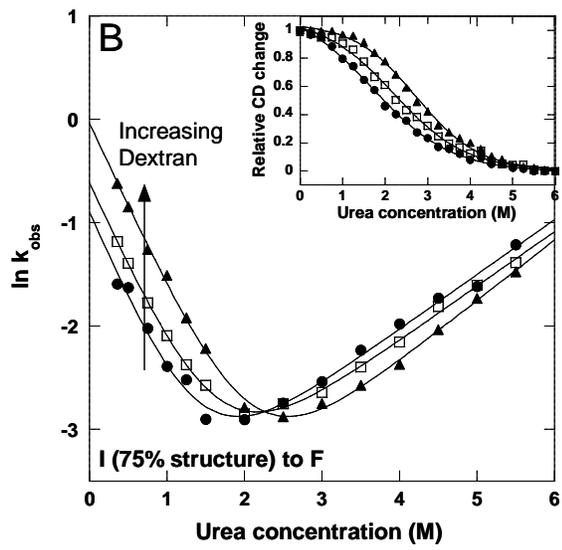